\documentclass[conference]{IEEEtran}
\IEEEoverridecommandlockouts
\usepackage{cite}
\usepackage{amsmath,amssymb,amsfonts}
\usepackage{pifont}
\usepackage{mathtools}
\usepackage{makecell}
\DeclareMathOperator*{\argmin}{arg\,min}

\usepackage{siunitx}

\usepackage[most]{tcolorbox}
\tcbset{boxrule=0.5pt,arc=2pt}

\usepackage{amsthm}
\newtheoremstyle{boldsmallcaps}
  {\topsep}                       
  {\topsep}                       
  {\itshape}                      
  {}                              
  {\bfseries\scshape}             
  {.}                             
  { }                             
  {}                              
\theoremstyle{boldsmallcaps} 
\newtheorem{remark}{Remark}

\usepackage{algorithmic}
\usepackage[ruled,vlined]{algorithm2e}
\usepackage{enumitem} 
\usepackage{graphicx}
\usepackage{textcomp}
\usepackage{float}
\usepackage[colorlinks=true, linkcolor=blue, citecolor=blue, urlcolor=blue]{hyperref}
\usepackage{xcolor}
\def\BibTeX{{\rm B\kern-.05em{\sc i\kern-.025em b}\kern-.08em
    T\kern-.1667em\lower.7ex\hbox{E}\kern-.125emX}}
\usepackage{caption}
\usepackage{subcaption}
\usepackage{amsthm}

\usepackage{amssymb,amsfonts}
\usepackage{bm}
\usepackage{algorithmic}
\usepackage{graphicx}
\usepackage{textcomp}
\usepackage{color}  
\usepackage{booktabs}
\usepackage{caption}

\usepackage{footnote} 
\makesavenoteenv{table*} 

\usepackage[dvipsnames]{xcolor}

\newcommand{\helen}[1]{\textcolor{Orchid}{[{\em Helen}: #1]}}

\usepackage{amsmath}
 \renewcommand{\boldsymbol}{\mathbf}

\newcommand{\cp}[1]{\ifmmode {\mathcal{#1}}\else ${\mathcal{#1}}$\fi}
\usepackage{array}
\newcolumntype{P}[1]{>{\centering\arraybackslash}p{#1}}

\newcommand{\bx}{\boldsymbol{x}}

\usepackage{nccmath}
\def\cred{\textcolor{red}}

\definecolor{darkgreen}{rgb}{0., 0.4, 0.}
\definecolor{amber}{rgb}{1.0, 0.49, 0.0}
\definecolor{orange}{rgb}{1.0, 0.4, 0.0}

\newcommand{\tcospa}{TC-OSPA\textsuperscript{(2)} }
\newcommand{\ospasq}{OSPA\textsuperscript{(2)} }

\usepackage{multirow}
\usepackage{array} 
\usepackage[toc,acronyms,nopostdot,nogroupskip, sort=use, nonumberlist]{glossaries-extra} 
\MFUhyphentrue 
\glssetcategoryattribute{general}{glossdesc}{firstuc}
\glssetcategoryattribute{abbreviation}{glossdesc}{title}

\makeatletter
\newglossarystyle{long-initcapsdesc}{%
  \setglossarystyle{long}%
    \renewcommand{\glossentry}[2]{%
    \glsentryitem{##1}\glstarget{##1}{\glossentryname{##1}} &
    \protected@edef\thisdesc{\glsentrydesc{##1}}%
    \xcapitalisewords{\thisdesc}\glspostdescription\space ##2\tabularnewline
  }%
}
\makeatother
\makeglossaries

\usepackage{ragged2e}

\usepackage{comment}

\begin{document}
%
%
\newabbreviation{BSS}{BSS}{blind source separation}
\newabbreviation{RL}{RL}{reinforcement learning}
\newabbreviation{RJDM}{RJDM}{radar jamming decision-making}
\newabbreviation{LORO}{LORO}{lobe-on-receive-only}
\newabbreviation{MAP}{MAP}{maximum a posteriori}
\newabbreviation{MPC}{MPC}{model predictive control}
\newabbreviation{RPIP}{RPIP}{random pulse initial phases}
\newabbreviation{CP}{CP}{correlated processing}
\newabbreviation{CPI}{CPI}{coherent processing interval}
\newabbreviation{IC}{IC}{interference cancellation}
\newabbreviation{MUSIC}{MUSIC}{multiple signal classification}
\newabbreviation{SAM}{SAM}{swept amplitude-modulation}
\newabbreviation{AM}{AM}{swept amplitude-modulation}
\newabbreviation{NP}{NP}{Neyman-Pearson}
\newabbreviation{DBS}{DBS}{Doppler beam sharpening}
\newabbreviation{MTT}{MTT}{multiple target tracking}
\newabbreviation{MOT}{MOT}{multi-object tracking}
\newabbreviation{MOF}{MOF}{multi-object filtering}
\newabbreviation{DMOT}{DMOT}{distributed MOT}
\newabbreviation{LFM}{LFM}{linear frequency-modulated}
\newabbreviation{LOS}{LOS}{line-of-sight}
\newabbreviation[plural={directions of arrival}]{DoA}{DoA}{direction of arrival}
\newabbreviation{SPJ}{SPJ}{self-protection jamming}
\newabbreviation{SOJ}{SOJ}{stand-off jamming}
\newabbreviation{MLE}{MLE}{maximum likelihood estimator}
\newabbreviation[plural={neural networks}]{NN}{NN}{neural network}
\newabbreviation{EJ}{EJ}{escort jamming}
\newabbreviation{KL}{KL}{Kullback–Leibler}
\newabbreviation{TC}{TC}{track consensus}
\newabbreviation{LMB}{LMB}{labeled multi-Bernoulli }
\newabbreviation{MSA}{MSA}{measurement-space attack}  
\newabbreviation{SSA}{SSA}{state-space attack} 
\newabbreviation{SNA}{SNA}{social network analysis}
\newabbreviation{GDM}{GDM}{group decision-making}
\newabbreviation{OSPA}{OSPA}{optimal sub-pattern assignment}
\newabbreviation{GCI}{GCI}{generalized covariance intersection}
%
%
\newabbreviation{ITA}{ITA}{identity theft attack}
\newabbreviation{NITA}{NITA}{non-identity theft attack}
\newabbreviation{FDI}{FDI}{false data injection}
\newabbreviation{DoS}{DoS}{denial of service}
\newabbreviation{EPI}{EPI}{extra packet injection}
\newabbreviation{PHD}{PHD}{probability hypothesis density}
\newabbreviation{JRC}{JRC}{joint radar-communication}
\newabbreviation{WLS}{WLS}{weighted least squares}
\newabbreviation{ERP}{ERP}{effective radiated power}
\newabbreviation{LTI}{LTI}{linear time-invariant}
\newabbreviation{FoV}{FoV}{field of view}
\newabbreviation[shortplural={KFs}, plural={Kalman filters}]{KF}{KF}{Kalman filter}
\newabbreviation[shortplural={EKFs}, plural={extended Kalman filters}]{EKF}{EKF}{extended Kalman filter}
\newabbreviation[shortplural={PFs}, plural={particle filters}]{PF}{PF}{particle filter}
\newabbreviation[shortplural={UKFs}, plural={unscented Kalman filters}]{UKF}{UKF}{unscented Kalman filter}
\newabbreviation[shortplural={GSFs}, plural={Gaussian sum filters}]{GSF}{GSF}{Gaussian sum filter}
\newabbreviation{SLAM}{SLAM}{simultaneous localization and mapping}
\newabbreviation[shortplural={DOFs}, plural={degrees of freedom}]{DOF}{DOF}{degree of freedom}
\newabbreviation[shortplural={MMFTs}, plural={micro-motion false targets}]{MMFT}{MMFT}{micro-motion false target}
\newabbreviation[shortplural={TFTs}, plural={translational false targets}]{TFT}{TFT}{translational false target}
\newabbreviation[shortplural={PRIs}, plural={pulse repetition interval}]{PRI}{PRI}{pulse repetition interval}
\newabbreviation[shortplural={PTs}, plural={physical targets}]{PT}{PT}{physical target}
\newabbreviation{FFT}{FFT}{fast Fourier transform}
\newabbreviation{SSM}{SSM}{state-space model}
\newabbreviation{JSR}{JSR}{Jamming-to-Signal Ratio}
\newabbreviation{RCS}{RCS}{radar DAcross-section}
\newabbreviation{CS}{CS}{compressed sensing}
\newabbreviation{AI}{AI}{artificial intelligence}
\newabbreviation{EW}{EW}{electronic warfare}
\newabbreviation{CW}{CW}{continuous-wave}
\newabbreviation{IF}{IF}{intermediate frequency}
\newabbreviation[shortplural={TOIs}, plural={targets of interest}]{TOI}{TOI}{target \MakeLowercase{o}f interest}
\newabbreviation[shortplural={ECMs}, plural={electronic countermeasures}]{ECM}{ECM}{electronic countermeasure}
\newabbreviation[shortplural={ECCMs}, plural={electronic counter-countermeasures}]{ECCM}{ECCM}{electronic counter-countermeasure}
\newabbreviation[shortplural={FTs}, plural={false targets}]{FT}{FT}{false target}
\newabbreviation[shortplural={PDs}, plural={pulse dopplers}]{PD}{PD}{pulse doppler}
\newabbreviation[shortplural={FTGs}, plural={false target generator}]{FTG}{FTG}{false target generator}
\newabbreviation[shortplural={FDAs}, plural={frequency diverse arrays}]{FDA}{FDA}{frequency diverse array}
\newabbreviation[shortplural={SARs}, plural={synthetic aperture radars}]{SAR}{SAR}{synthetic aperture radar}
\newabbreviation[shortplural={DRFMs}, plural={digital radio-frequency memories}]{DRFM}{DRFM}{digital radio frequency memory}
\newabbreviation[shortplural={RGPOs}, plural={range gate pull-offs}]{RGPO}{RGPO}{range gate pull-off}
\newabbreviation[shortplural={RGPIs}, plural={range gate pull-ins}]{RGPI}{RGPI}{range gate pull-in}
\newabbreviation[shortplural={VGPOs}, plural={velocity gate pull-offs}]{VGPO}{VGPO}{velocity gate pull-off}
\newabbreviation[shortplural={VGPI}s, plural={velocity gate pull-ins}]{VGPI}{VGPI}{velocity gate pull-in}
\newabbreviation[shortplural={RVGPOs}, plural={range-velocity gate pull-offs}]{RVGPO}{RVGPO}{range-velocity gate pull-off}
\newabbreviation[shortplural={RVGPIs}, plural={range-velocity gate pull-ins}]{RVGPI}{RVGPI}{range-velocity gate pull-in}
\newabbreviation[shortplural={ISRJs}, plural={interrupted-sampling repeater jammings}]{ISRJ}{ISRJ}{interrupted-sampling repeater jamming}
\newabbreviation{CRDJ}{CRDJ}{crosspulse repeater deception jamming}
\newabbreviation[shortplural={MHTs}, plural={multiple hypothesis trackings}]{MHT}{MHT}{multiple hypothesis tracking}
\newabbreviation[shortplural={RNNs}, plural={recurrent neural networks}]{RNN}{RNN}{recurrent neural network}
\newabbreviation[shortplural={SJNRs}, plural={Signal-to-Jammer Noise Ratios}]{SJNR}{SJNR}{Signal-to-Jammer Noise Ratio}
\newabbreviation[shortplural={CNNs}, plural={convolutional neural networks}]{CNN}{CNN}{convolutional neural network}
\newabbreviation[shortplural={LSTMs}, plural={long short-term memories}]{LSTM}{LSTM}{long short-term memory}
\newabbreviation[shortplural={TDOAs}, plural={time differences of arrival}]{TDOA}{TDOA}{time difference of arrival}
\newabbreviation[shortplural={RFSs}, plural={random finite sets}]{RFS}{RFS}{random finite set}
\newabbreviation[shortplural={SNRs}, plural={signal-to-noise ratios}]{SNR}{SNR}{signal-to-noise ratio}
\newabbreviation[shortplural={SJRs}, plural={signal-to-jammer ratios}]{SJR}{SJR}{signal-to-jammer ratio}
\newabbreviation[shortplural={OFDMs}, plural={orthogonal frequency-division multiplexings}]{OFDM}{OFDM}{orthogonal frequency-division multiplexing}
\newabbreviation[shortplural={PRFs}, plural={pulse repetition frequencies}]{PRF}{PRF}{pulse repetition frequency}
\newabbreviation[shortplural={SIMOs}, plural={single-input multiple-outputs}]{SIMO}{SIMO}{single-input multiple-output}
\newabbreviation{DA}{DA}{data association}
\newabbreviation[shortplural={MIMOs}, plural={multiple-input multiple-outputs}]{MIMO}{MIMO}{multiple-input multiple-output}
\newabbreviation[shortplural={GLRTs}, plural={generalized likelihood ratio tests}]{GLRT}{GLRT}{generalized likelihood ratio test}
\newabbreviation[shortplural={AGCs}, plural={automatic gain controls}]{AGC}{AGC}{automatic gain control}
\newabbreviation[shortplural={MLs}, plural={machine learnings}]{ML}{ML}{machine learning}
\newabbreviation[shortplural={UAVs}, plural={unmanned aerial vehicles}]{UAV}{UAV}{unmanned aerial vehicle}
\newabbreviation[shortplural={GNSSs}, plural={global navigation satellite systems}]{GNSS}{GNSS}{global navigation satellite systems}

%
\title{Label Hijacking in Track Consensus-Based Distributed Multi-Target Tracking
\thanks{This work has been partially supported by the National Science Foundation under Awards 1845833, 2326559 and 2530870.}
}

\author{
    Helena Calatrava, Shuo Tang, Pau Closas\\
    \textit{Northeastern University, Boston, MA 02115, USA} \\
    \{calatrava.h, tang.shu, closas\}@northeastern.edu
}

\maketitle

\begin{abstract}
Distributed multi-target tracking (DMTT) in limited field-of-view (FoV) sensor networks commonly suffers from \textit{label inconsistency}, whereby different nodes disagree on the identity of the same target.
Recent track-consensus DMTT (TC-DMTT) strategies mitigate this issue by enforcing kinematic and label agreement through metric-based track matching. Nevertheless, their behavior under adversarial conditions remains largely unexplored.
%
%
In this paper, we reveal identity-level vulnerabilities in TC-DMTT and introduce the concept of \textit{label hijacking}: an attack in which an adversary injects spoofed tracks to corrupt target identities across the network.
Drawing on an analogy to classical pull-off deception in radar, we formalize a notion of attack stealthiness and derive an optimization-based strategy for crafting such attacks.
%
%
A three-sensor network case study demonstrates the impact of the proposed attack on label consistency and tracking accuracy, showing successful target impersonation.
Overall, this work highlights the need to rethink robustness at the consensus layer in DMTT frameworks.

\end{abstract}


\section{Introduction}\label{sec:intro}
%



\Gls{MTT} is a statistical inference problem concerned with estimating the number of physical targets present in a dynamic scene, their trajectories, and, in certain formulations, their \textit{identities (labels)} from noisy sensor measurements~\cite{bar2011tracking}. Originating in the 1960s, \gls{MTT} has since become a foundational technology for a wide range of perception and decision-making applications~\cite{vo2015multitarget}.




Modern \gls{MTT} systems must contend with multiple sources of uncertainty beyond process and measurement noise, including time-varying target cardinality, false alarms (clutter), missed detections, target births and deaths, and \gls{DA}.
Inspired by Mahler’s work on finite set statistics~\cite{mahler_statistical_2007}, \gls{RFS} theory emerged as a mathematically elegant framework for handling these uncertainties implicitly within a unified Bayesian formulation~\cite{vo2008random, wu2023target}.
%
%
%

A fundamental question in \gls{MTT} concerns how information should be processed in \textit{multi-sensor settings}, where transmitting all raw measurements to a central processor is often impractical~\cite{souza2016target}. To address this, \textit{distributed MTT (DMTT)} allows each node to perform local tracking and exchange summary information for fusion, either in the form of track-level state estimates (commonly referred to as \textit{track-to-track fusion}) or multi-object densities. Because the \gls{RFS} framework naturally generalizes single-object distributed estimation to the multi-object case~\cite{mahler2000optimal}, a wide range of \gls{RFS}-based filters, including labeled formulations~\cite{fantacci2018robust}, have been adapted to the distributed setting, predominantly relying on \gls{GCI}~\cite{hurley2002information} for log-linear geometric averaging of probability densities.



\begin{figure}[t]
\centering
  \includegraphics[width=.85\columnwidth]{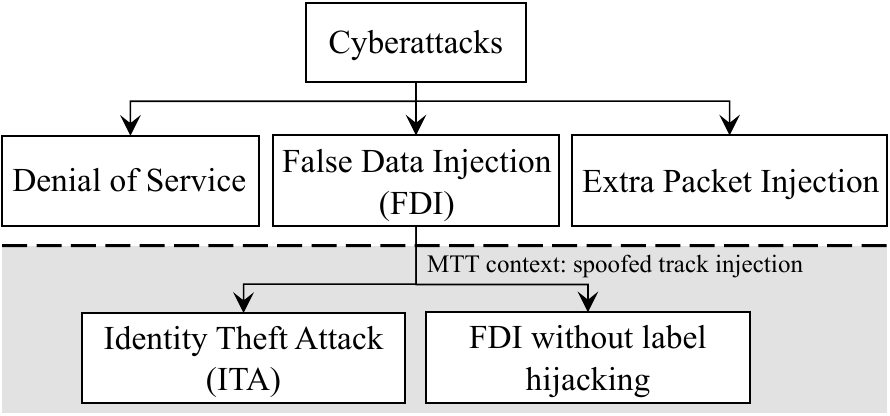}
  \caption{Cyberattack taxonomy for distributed sensor networks~\cite{yu2021secure}. We specialize false data injection to the MTT setting, focusing on identity threats.
}  
  \label{fig:intro:attack_classification}
  \vspace{-.5cm}
\end{figure}

%
%

%

While widely adopted, \gls{GCI}-based fusion methods suffer from two key limitations. First, under \textit{limited and non-overlapping \gls{FoV}} conditions, their performance degrades~\cite{uney2019fusion}, and labeled formulations are prone to \textit{label inconsistency}, whereby different nodes may assign distinct identities to the same physical target~\cite{li2017robust}. Second, more generally, density-level fusion incurs significant computational and communication overhead compared to track-to-track approaches~\cite{8753710}.
Recent work on \textit{track-consensus} DMTT (TC-DMTT)~\cite{van2021distributed,verma2023track} (reviewed in Sec.~\ref{sec:bckground}) addresses these challenges through enhanced track-to-track fusion, enforcing both kinematic and label consensus using track-distance metrics, typically drawn from the \gls{OSPA} family~\cite{8217598}.
While effective under benign operating conditions, TC-DMTT performance in the presence of adversarial threats has received limited attention.

\begin{figure*}[t]
\centering
\fbox{\includegraphics[width=.85\textwidth]{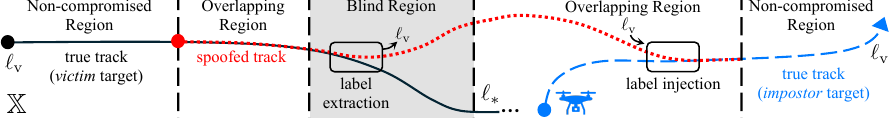}}
  \caption{\textcolor{black}{We introduce \textit{label hijacking (extraction + injection)} as the data-layer analogue of pull-off radar deception~\cite{Calatrava2024DeceptionSurvey}.} The spoofed track hijacks label $\ell_\text{v}$ from the \textit{victim}
  target, which is then tracked under an incorrect label $\ell_\ast$. After hijacking $\ell_\text{v}$, the spoofed track can move freely while remaining associated with that identity; by aligning with another target, $\ell_\text{v}$ can be transferred to an \textit{impostor} target.}  
   \label{fig:intro:motivation}
  \vspace{-.5cm}
\end{figure*}

Although distributed architectures improve observability under limited \gls{FoV}, they simultaneously expand the attack surface to cyberattacks~\cite{yu2021secure} in distributed networks (see Fig.~\ref{fig:intro:attack_classification} for a classification).
In this work, we focus on a false data injection (FDI) scenario in which an attacker injects \textit{spoofed tracks}\footnote{Distinct from \textit{false tracks} in the MTT literature, which arise from false alarms or data association errors.}, i.e., tracks that do not correspond to any true (physical) target, into the fusion process.
Such attacks may arise when an adversary compromises a sensor’s communication channel and manipulates the information disseminated to other, non-compromised nodes. This renders the affected node unreliable and corrupts the distributed fusion process.

%

%
%
%
%
Provided that true targets remain accurately tracked, the mere presence of spoofed tracks may be tolerable in some applications. Nevertheless, a more critical failure mode arises when target identities are corrupted.
As illustrated in Fig.~\ref{fig:intro:motivation}, under \textit{label hijacking}, the fusion process continues to follow a true target under incorrect labels, giving rise to what we term an \textit{\gls{ITA}} and breaking track continuity. \textcolor{black}{We further show that this threat model admits a principled parallel with deception jamming in radar systems~\cite{Calatrava2024DeceptionSurvey}.}
%


%
%


By enforcing label consensus through track-distance constraints, TC-DMTT architectures introduce identity-level vulnerabilities that remain largely unexplored.
In this paper, we show that the metric-based track-matching stage constitutes a critical attack surface enabling label hijacking.
Our main contributions are:
\textit{(i)} we introduce the track-consensus identity-theft attack (TC-ITA), which exploits this vulnerability through adversarial manipulation of track consensus;
\textit{(ii)} \textcolor{black}{by framing TC-ITA as the data-layer analogue of classical pull-off radar deception,} we define a notion of \textit{stealthiness} and propose an \gls{MPC}-based strategy to generate such attacks; and
\textit{(iii)} we demonstrate successful identity hijacking under stealthy TC-ITA in a simulated distributed tracking scenario\footnote{Code available at \url{https://github.com/hcalatrava/dmtt-adversarial}\label{refnote}}.

The remainder of this paper is organized as follows. Section~\ref{sec:bckground} reviews TC-DMTT frameworks; Section~\ref{sec:method} discusses adversarial opportunities against these frameworks and introduces the proposed attack model; Section~\ref{sec:results} presents the results; and Section~\ref{sec:conclusion} concludes the paper.

\section{Background: Track Consensus-Based Distributed Multi-Target Tracking}\label{sec:bckground}

This section reviews \emph{track-consensus} DMTT frameworks, which we collectively refer to as \emph{TC-DMTT}. Their distributed architecture and track-matching metrics are summarized in Secs.~\ref{sec:system_model} and \ref{sec:tcdmtt:ospa}, respectively. TC-DMTT extends classical track-to-track fusion to resolve label inconsistencies in networks with limited and non-overlapping \glspl{FoV} through track consensus, as reviewed in Sec.~\ref{sec:tc}.

\textbf{Notation:} Bold symbols denote labeled quantities; uppercase symbols denote set-valued quantities. Our notation largely follows the work in~\cite{van2021distributed}.

\subsection{System Model}\label{sec:system_model}
%

Fusion is performed in a fully distributed manner, without global knowledge of the network topology or FoV of other nodes.
Moreover, because the fusion stage operates only on the reported track
estimates and does not modify the local multi-object densities,
TC-DMTT is agnostic to the specific local MTT algorithms
employed at each node.

\subsubsection{Network topology}
%
%
A distributed sensor network is modeled as an undirected graph
\( G = (\mathcal{N}, \mathcal{A}) \), where \( \mathcal{N} \) denotes the set of nodes and
\( \mathcal{A} \subseteq \mathcal{N} \times \mathcal{N} \) the set of bidirectional
communication links. Two nodes \( a,b \in \mathcal{N} \) can communicate directly
if they are within communication range~\cite{yu2021secure}. The neighbor set of node \( a \) is defined
as $\mathcal{N}^{(a)} \triangleq \{ b \in \mathcal{N} \mid b \neq a \text{ and } (a,b) \in \mathcal{A} \}$.
Each node is equipped with a limited \gls{FoV} sensor subject to false alarms and
detection uncertainty, as well as local processing capabilities.

%

\subsubsection{Message exchange and global labeling}
At time \( k \), each node \( a \) transmits its local labeled multi-object state estimate
\( \mathbf{X}_k \subset \mathbb{X} \times \mathbb{L} \) to the other nodes, where each element
\( \bx_k = (x_k, \ell_k) \) consists of a kinematic state $x_k \in \mathbb{X}$ and a locally unique label $\ell_k \in \mathbb{L}$.
Since local labels are not comparable across nodes, a global labeling scheme is obtained by augmenting each label with the identity of the transmitting node, yielding
\( (\ell_k, a) \in \mathbb{I} = \mathbb{L} \times \mathcal{N} \).
Each node stores the globally labeled multi-object state estimates received from each transmitting neighbor \(a\), denoted by
\( \mathbf{X}^{(a)}_k \subset \mathbb{X} \times \mathbb{I} \). 

\subsubsection{Track formation}
A \textit{track} is a time-ordered sequence of single-object state estimates over the subset of time indices
\( \mathcal{D}^{(t)} \subseteq \mathcal{K} \) at which the object is declared to exist.
Each node \(a\) independently maintains a local track memory for its own
estimates \(\mathbf{X}^{(a)}_k\) and for those received from its neighbors
\(\{\mathbf{X}^{(b)}_k\}_{b\in\mathcal{N}(a)}\), over a finite time window
\(\mathcal{K}\triangleq\{1,\dots,K\}\).
For each source \(b\in\mathcal{N}(a)\cup\{a\}\), a corresponding set of
globally labeled tracks \(\mathbf{T}^{(b)}\subset\mathbb{T}\times\mathbb{I}\)
is constructed by linking states that share the same global label across time.

\subsection{Track Consensus}\label{sec:tc}
%

Track consensus in TC-DMTT consists of two stages: \textit{(1)} kinematic consensus, which aligns and fuses track states, and \textit{(2)} label consensus, which propagates identity agreements across the network.

\subsubsection{Kinematic consensus}\label{sec:kin_consensus}
\begin{figure}[t]
\centering
  \includegraphics[width=.85\columnwidth]{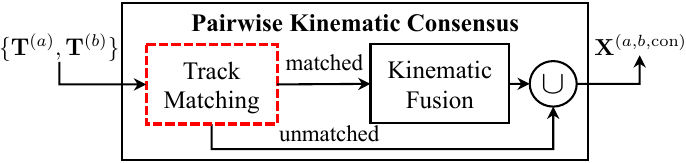}
\caption{Pairwise kinematic consensus between two nodes.
Tracks in $\{\mathbf{T}^{(a)},\mathbf{T}^{(b)}\}$ are matched using a
track-distance metric (Sec.~\ref{sec:tcdmtt:ospa}); matched tracks are fused
and combined with retained unmatched tracks to form
$\mathbf{X}^{(a,b,\mathrm{con})}$. As shown in Sec.~\ref{sec:method}, manipulating the track-matching stage alone is
sufficient to induce \emph{identity theft} (see Remark~\ref{remark:attack}).
}
 \label{fig:tc_diagram}
  \vspace{-.5cm}
\end{figure}

At time \(k\), node \(a\) has access to its own active track set \(\mathbf{T}^{(a)}\)
and those from its neighbors, i.e.,
\(\{\mathbf{T}^{(b)}\}_{b\in\mathcal N(a)}\). Here, \textit{active} refers to tracks declared to exist and whose state estimates are available for fusion.
Pairwise kinematic consensus is then performed using the two-node procedure
in Fig.~\ref{fig:tc_diagram}: for each neighbor \(b\in\mathcal N(a)\),
node \(a\) applies this operation to the pair
\(\{\mathbf{T}^{(a)},\mathbf{T}^{(b)}\}\),
thereby producing a collection of intermediate consensus results \(\{\mathbf{X}^{(a,b,\text{con})}\}_{b\in\mathcal{N}(a)}\).
These intermediate outputs are subsequently combined sequentially to form
the full (network) kinematic consensus at node \(a\), denoted as $\mathbf{X}^{a,\text{con}}$ (see~\cite{van2021distributed, verma2023track} for details).

\smallskip
\noindent We now describe the two-node (pairwise) kinematic consensus procedure
between the track sets $\mathbf{T}^{(a)}$ and $\mathbf{T}^{(b)}$ in detail,
which consists of a track-matching stage followed by kinematic fusion
(Fig.~\ref{fig:tc_diagram}).
The role of track matching is to answer the question:
\emph{Which track in $\mathbf{T}^{(a)}$ corresponds to which track in
$\mathbf{T}^{(b)}$?}
First, an optimal assignment between the two track sets is computed using a
track-distance metric, yielding a one-to-one matching between tracks.
We review the most common choice of track metric in Sec.~\ref{sec:tcdmtt:ospa}.
The resulting pairs are then filtered using a cut-off distance $c$ (i.e., only pairs with cost below $c$ are considered valid matches).
Each matched pair is subsequently fused to produce a single track estimate,
typically via a weighted combination of the individual state estimates,
where the weights reflect the relative confidence of each node.
Unmatched tracks may be retained according to heuristic criteria
(e.g., minimum track length)~\cite{van2021distributed,verma2023track}.

\subsubsection{Label consensus}
Matched tracks identified during the pairwise kinematic-consensus stage
(Fig.~\ref{fig:tc_diagram}) are treated as local identity agreements, and are
propagated transitively through the network via a label-consensus mechanism
\cite{van2021distributed,verma2023track}.
We omit the specifics of this stage, as the proposed attack induces identity corruption through compromised track matching, which is then propagated by the label consensus mechanism (see Remark~\ref{remark:attack}).

\subsection{Track Matching Metric}\label{sec:tcdmtt:ospa}
%
%
Kinematic consensus requires computing distances between track sets, which in turn
necessitates an appropriate track-matching metric.
%
%
%
The works in~\cite{verma2023track,van2021distributed} adopt for this purpose the
temporal extension of the \gls{OSPA} metric, known as \textit{OSPA-on-OSPA} and
denoted \gls{OSPA}$^{(2)}$.
While we define both metrics below, a detailed discussion of their properties
can be found in~\cite{8217598}.

%
%

\subsubsection{OSPA}
Between two finite sets \( X, Y \in \mathcal{F}(\mathbb{X}) \), where \( \mathcal{F}(\mathbb{X}) \) is the space of finite sets of \( \mathbb{X} \), $X = \{x^{(1)}, \dots, x^{(m)}\}$ and $Y = \{y^{(1)}, \dots, y^{(n)}\}$, the $p$-order OSPA is defined as
\begin{equation}\label{eq:ospa}
\begin{split}
    &d_{p}^{(c)}(X,Y) \\&=
    \left( 
    \frac{1}{n} \left( 
    \min_{\pi \in \Pi_n} \sum_{i=1}^{m} d^{(c)}(x^{(i)}, y^{(\pi(i))})^p + c^p (n - m) 
    \right) 
    \right)^{1/p}
    \end{split}
\end{equation}
for \( m \leq n \), and \( d_{p}^{(c)}(X,Y) \triangleq d_{p}^{(c)}(Y,X) \) for \( m > n \). 
Here, \( \Pi_n \) denotes the set of all permutations of \( \{1,\dots,n\} \); \( d^{(c)}(x,y) = \min(c,d(x,y)) \) denotes the truncated base distance, where $c\in(0,\infty)$ is the cut-off parameter; and \( d(\cdot,\cdot) \) is a metric on \( \mathbb{X} \), e.g., the Manhattan distance. Note that OSPA is defined with the conventions \( d_{p}^{(c)}(\emptyset, \emptyset) = 0 \), and \( d_{p}^{(c)}(X,Y) = c \) if one argument is empty, whereby all distances exceeding \(c\) are treated as equally large.


\subsubsection{OSPA$^{(2)}$}
This metric extends \gls{OSPA} to the space of finite sets of tracks \( \mathbb{T} \), i.e., \( \mathcal{F}(\mathbb{T}) \), as a time-averaged OSPA distance. For tracks \( t, u \in \mathbb{T} \), \ospasq is given by

\begin{equation}
\begin{aligned}\label{eq:ospa_track2track}
    \hspace{-.3cm} &\tilde{d}^{(c)}(t,u) \\&=
    \begin{cases} 
    \sum\limits_{k \in \mathcal{D}^{(t)} \cup \mathcal{D}^{(u)}}
    \frac{d^{(c)}(\{t(k)\}, \{u(k)\})}{|\mathcal{D}^{(t)} \cup \mathcal{D}^{(u)}|}, & \text{if } \mathcal{D}^{(t)} \cup \mathcal{D}^{(u)} \neq \emptyset, \\
    0, &  \text{otherwise}.
    \end{cases}
    \end{aligned}
    \end{equation}
%
%
where \( d^{(c)}(\{t(k)\}, \{u(k)\}) \) is the \gls{OSPA} distance between tracks evaluated at time instant \( k \), i.e., between singleton sets, which enforces $p=1$.
Here, we denote by \( t(k) \in \mathbb{X} \) the single-object state of track \( t \) at time \( k \in \mathcal{D}^{(t)} \).

\section{Adversarial Model}\label{sec:method}
This section presents the adversarial model and feasibility assumptions for label hijacking (Sec.~\ref{sec:attack:scenario}), introduces the proposed TC-ITA strategy (Sec.~\ref{sec:attack_description}), \textcolor{black}{and develops an \gls{MPC}-based approach for generating stealthy spoofed trajectories inspired by the takeover effect in RGPO attacks (Sec.~\ref{sec:attack:mpc})~\cite{Calatrava2024DeceptionSurvey}.}

The \textit{attack goal} is to transfer the global label of a \textit{victim target} to a controlled \textit{impostor target}, such that the network continues to believe it is tracking the victim while in reality following the impostor. Remarkably, if the impostor is an attacker-controlled agent (e.g., a drone), its motion can be arbitrarily designed to induce any desired behavior under the stolen identity.

\textbf{Terminology:} A \textit{compromised node} is a sensor node whose communication channel is controlled by the attacker. A \textit{non-compromised node} is a sensor node that is not under adversarial control; however, the attacker may intercept its transmitted messages or seek to influence its fusion output. A \textit{blind region} is a region of the surveillance area that is not within the \gls{FoV} of any non-compromised node. A \textit{non-compromised region} is a region observed by at least one non-compromised node.

\subsection{Adversarial Scenario}\label{sec:attack:scenario}

\begin{figure}[t]
\centering
  \includegraphics[width=.8\columnwidth]{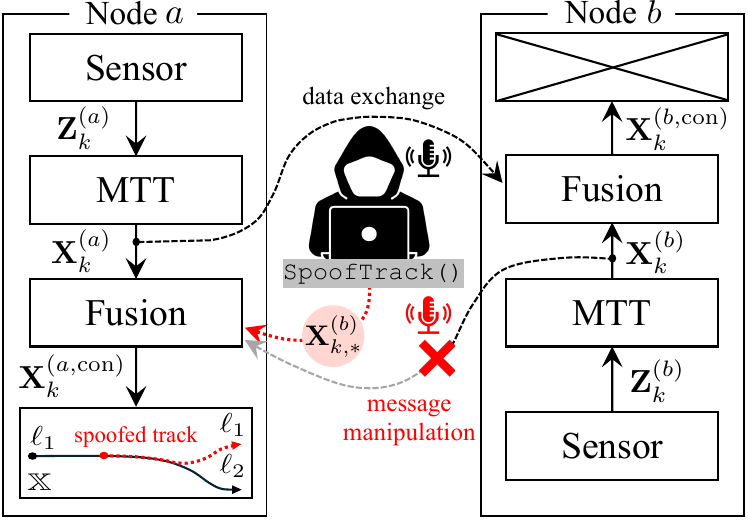}
\caption{False-data injection scenario in TC-DMTT (naturally extensible to larger
networks). At time $k$, each node $i$ forms local tracks
$\mathbf{X}_k^{(i)}$ from its measurements $\mathbf{Z}_k^{(i)}$.
The attacker compromises node $b$, replaces its nominal tracks
$\mathbf{X}_k^{(b)}$ with adversarial ones $\mathbf{X}_{k,\ast}^{(b)}$, and
forwards them to non-compromised neighbors, thereby corrupting their local
fusion outputs $\mathbf{X}_k^{(a,\mathrm{con})}$.
}
 \label{fig:scheme_attack}
\end{figure}

We consider the adversarial scenario illustrated in Fig.~\ref{fig:scheme_attack}, in which an attacker compromises one or more sensor nodes by gaining control over their communication channels.
The attacker intercepts both the local track estimates produced at the compromised nodes and those transmitted by their non-compromised neighbors. By aggregating this information, the attacker gains access to target trajectories even outside the sensing coverage of the compromised nodes, as long as the target is observed by at least one neighbor.

From the intercepted data, the attacker constructs spoofed tracks and forwards them in place of the nominal local track estimates of the compromised nodes, thereby corrupting the local fusion processes and resulting consensus estimates of neighboring nodes.
This mechanism follows the same intercept–modify–reinject paradigm as a \textit{replay attack}.
%




%
The proposed attack is developed under the following assumptions:
\begin{enumerate}
    \item[A1)] Sensor nodes have limited and partially non-overlapping \glspl{FoV}. This allows the victim target to enter a blind region, which is required for label extraction.
    \item[A2)] The attacker must compromise direct neighbors of the nodes whose fusion processes it seeks to manipulate and of the nodes from which it aims to intercept track information.
    \item[A3)] Since nodes lack knowledge of the network geometry, non-compromised nodes cannot verify whether a reported track is physically observed by its sending node, and therefore accept spoofed tracks as valid during consensus.
    \item[A4)] The attacker can infer when the victim target enters and exits a blind region.
\end{enumerate}

%

The \emph{visibility inference} in Assumption~A4 enables the attacker to time the label extraction and injection stages described in Sec.~\ref{sec:attack_description}. It is obtained heuristically from intercepted communication patterns: the absence of reported updates over several time steps indicates that a target has left a node’s \gls{FoV}, whereas continued reporting implies ongoing observability. Remarkably, the delay in detecting these transitions, which we assume to be negligible, depends on the local MTT track-management policy rather than on any knowledge of the network geometry. 


\subsection{Identity Theft Attack Strategy}\label{sec:attack_description}
%
%

%
%

\begin{figure}[t]
\centering
\setlength{\fboxrule}{0.4pt} 
\fbox{\includegraphics[width=.85\columnwidth]{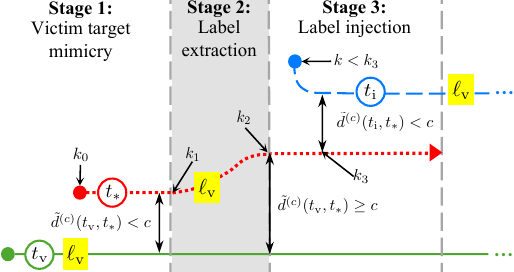}}
\caption{Overview of the three stages of TC-ITA and the associated notation (see Sec.~\ref{sec:attack_description}). 
White and gray areas denote non-compromised and blind regions, respectively.
}
\label{fig:attack}
\vspace{-.5cm}
\end{figure}


%
%

%

%

Under TC-ITA, the attacker selects a victim target tracked by the network with global label $\ell_{\mathrm{v}}$ and monitors its state estimates as reported by neighboring and compromised nodes.
Let $t_{\mathrm{v}}$ and $t_{\mathrm{i}}$ denote representative victim and impostor tracks, as reported by at least one node accessible to the attacker (compromised or non-compromised).
When multiple nodes report either target, the timing and track-distance constraints introduced below are assumed to hold for all corresponding tracks.
Additionally, let $t_\ast$ denote the spoofed track transmitted by the attacker starting at time $k_0$.
We now describe the three stages of TC-ITA, as depicted in Fig.~\ref{fig:attack}.

%



\subsubsection{Victim target mimicry}\label{sec:stage1}
$t_\ast$ assumes the victim’s global label $\ell_\text{v}$ by \emph{mimicking} $t_{\text{v}}$ closely enough to be selected by the track-matching rule in Sec.~\ref{sec:kin_consensus} (see Remark~\ref{remark:single_step_sufficiency}).
This requires satisfying the matching condition $\tilde d^{(c)}\!\left(t_{\text{v}}, t_\ast\right) < c$ for some $k \in [k_0, k_1)$, i.e., before the victim enters the blind region.
At time $k_1$, the attacker detects that the victim enters the blind region (Assumption~A4), which triggers the start of the label extraction stage.



%


\subsubsection{Label extraction \textcolor{black}{(pull-off stage)}}\label{sec:stage:lab_extraction}
%
%
%
In the blind region, i.e., for $k \in (k_1, k_2)$, the network receives only $t_\ast$ as the candidate track associated with label $\ell_{\text{v}}$. Since no competing track exists in this period, $t_\ast$ can evolve arbitrarily while retaining the hijacked label.
Upon the victim’s re-entry to the non-compromised region at time $k_2$, the separation condition $\tilde d^{(c)}\!\left(t_{\mathrm{v}}, t_\ast\right) \geq c$ must hold to ensure that $t_{\mathrm{v}}$ is considered a new object and therefore assigned a different label than $\ell_{\mathrm{v}}$ after confirmation; this condition must be maintained thereafter to prevent re-matching.
Note that if the victim target does not re-enter the non-compromised region, $t_\ast$ is not at risk of being matched with $t_{\text{v}}$, since $\tilde d^{(c)}\!\left(t_{\text{v}}, t_\ast\right) = \tilde d^{(c)}\!\left(\emptyset, t_\ast\right) =  c$.
Remarkably, this stage is directly analogous to a pull-off deception in radar tracking: whereas classical RGPO pulls the radar range gate away from the true target return, TC-ITA pulls off the target's identity label.
Note that label pull-off is achieved regardless of the stealthiness of $t_\ast$ during the blind interval (see the hard-switch variant in Sec.~\ref{sec:results}); the stealthy variant is further inspired by the RGPO takeover regime, providing kinematic plausibility by enforcing a smooth trajectory transition.
%

%

\subsubsection{Label injection}
%
%
Assuming that the impostor target has entered the non-compromised region, the matching condition $\tilde d^{(c)}\!\left(t_{\mathrm{i}}, t_\ast\right) < c$ must be enforced at some time $k_3 > k_2$, causing $t_{\mathrm{i}}$ and $t_\ast$ to be matched and thus share a global label.
Since the track-matching mechanism retains the label of the longer-lived track, and assuming that the impostor target enters the scene later than the victim, the hijacked label $\ell_{\mathrm{v}}$ is propagated to $t_{\mathrm{i}}$.
If the injection is successful, then for $k > k_3$ the attacker can stop reporting $t_\ast$, after which $t_{\mathrm{i}}$ becomes the sole candidate associated with $\ell_{\mathrm{v}}$ and can thereafter evolve arbitrarily while retaining that identity.

\begin{remark}[Targeted track consensus stage] \label{remark:attack}
While the goal of TC-ITA is to hijack the victim target's label $\ell_\text{v}$, the attack does not manipulate the label-consensus mechanism itself. 
Instead, it operates entirely at the kinematic level by injecting a spoofed track $t_\ast$ that satisfies the track-matching constraints. 
This causes $t_\ast$ to be repeatedly matched with the victim track $t_{\mathrm{v}}$, drift during blind intervals, and eventually match an impostor track $t_{\mathrm{i}}$. 
Because label consensus is driven by persistent track matching, corrupting the kinematic consensus alone is sufficient to induce identity transfer.
%
\end{remark}

\begin{remark}[Single-step sufficiency]\label{remark:single_step_sufficiency}
Due to the time-averaged definition in~\eqref{eq:ospa_track2track}, a valid match between two tracks is established as soon as they are within distance $c$ at \emph{any} single time instant, irrespective of the window length.
This property is exploited in the mimicry and label-injection stages of TC-ITA, where enforcing proximity for only one time step suffices to trigger track matching.
During the label-extraction stage, the same property becomes critical: even a brief re-approach between the spoofed and victim tracks would immediately re-enable matching, necessitating sustained separation.
\end{remark}

\subsection{Model Predictive Control for Stealthy Attack Design}\label{sec:attack:mpc}

\begin{figure*}[t]
    \centering
    \begin{subfigure}[b]{0.32\textwidth}
        \centering
        \includegraphics[width=\linewidth]{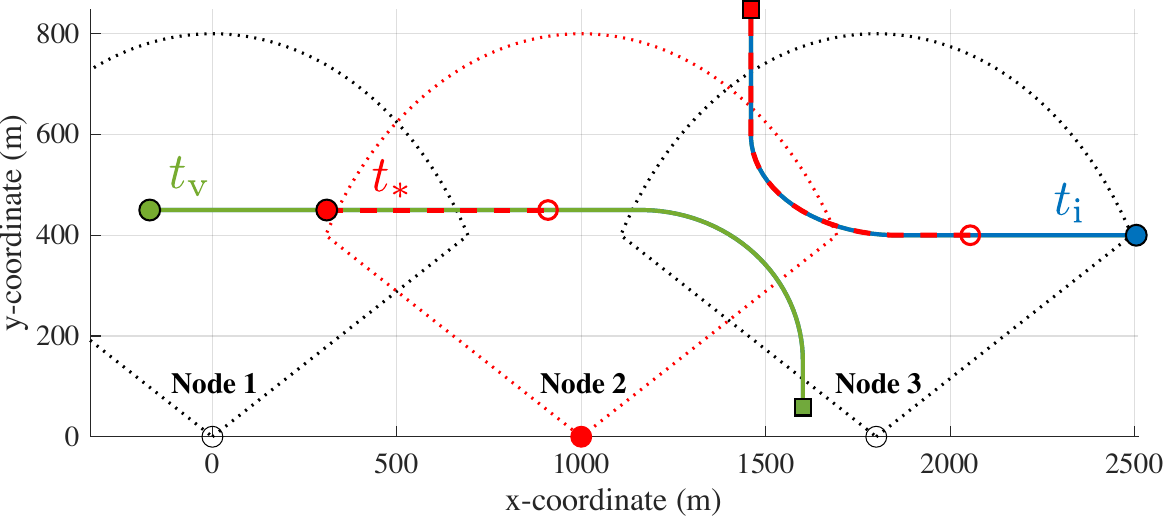}
        \caption{Ground truth (hard-switch).}\label{fig:results_1:gt_a1}
    \end{subfigure}
    \hfill
    \begin{subfigure}[b]{0.32\textwidth}
        \centering
        \includegraphics[width=\linewidth]{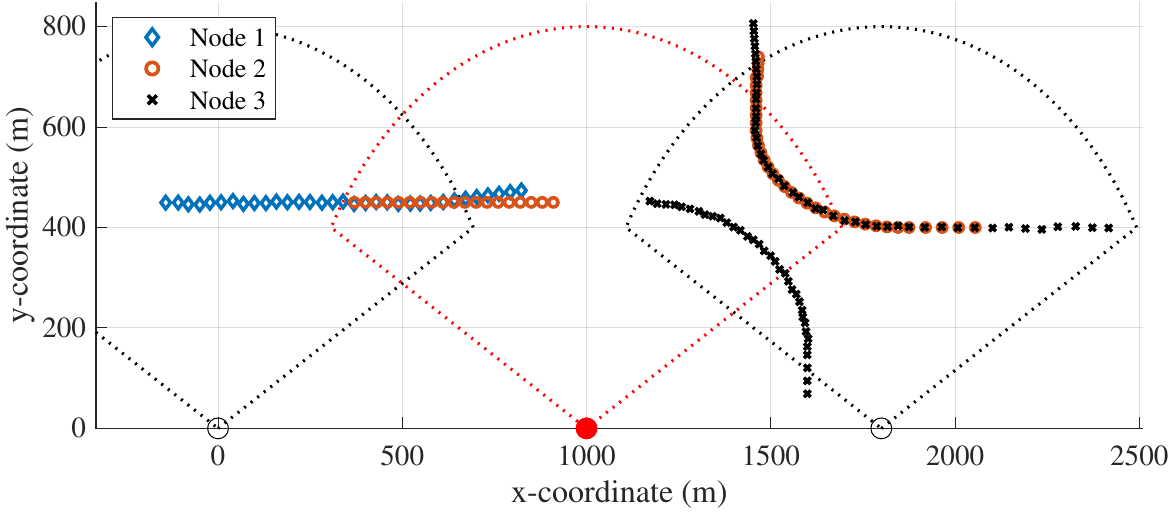}
        \caption{Individual estimates (hard-switch).}\label{fig:results_1:localEst_a1}
    \end{subfigure}
    \hfill
    \begin{subfigure}[b]{0.32\textwidth}
        \centering
        \includegraphics[width=\linewidth]{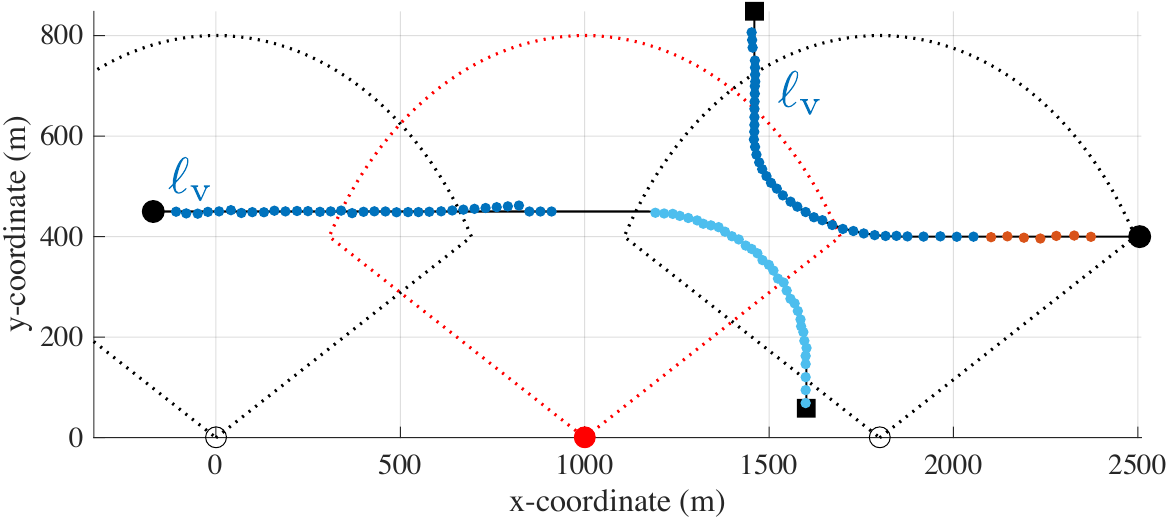}
        \caption{Consensus estimation (hard-switch).}\label{fig:results_1:consensus_a1}
    \end{subfigure}
    \begin{subfigure}[b]{0.32\textwidth}
        \centering
        \includegraphics[width=\linewidth]{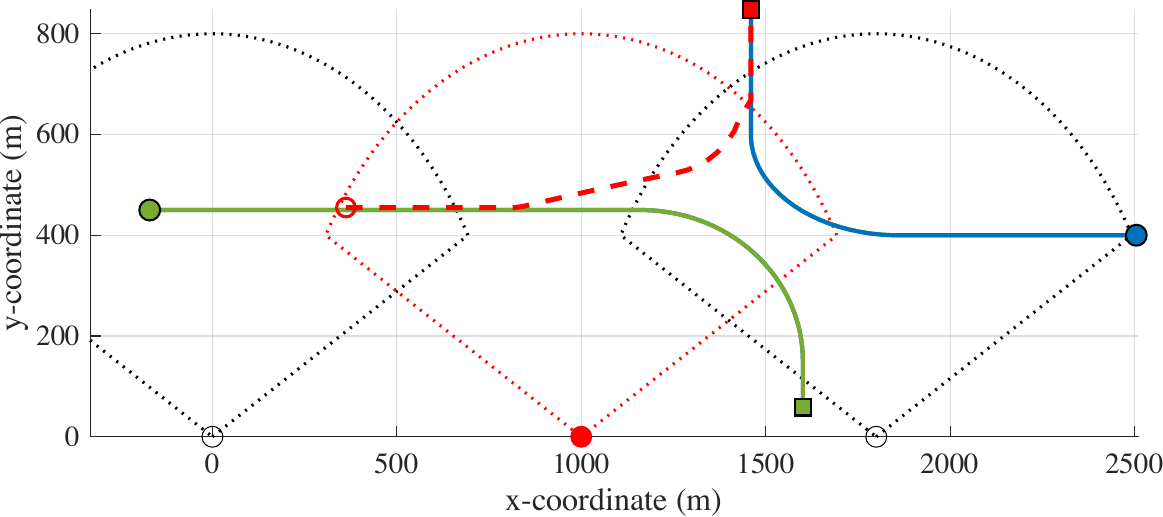}
        \caption{Ground truth (stealthy).}\label{fig:results_1:gt_a2}
    \end{subfigure}
    \hfill
    \begin{subfigure}[b]{0.32\textwidth}
        \centering
        \includegraphics[width=\linewidth]{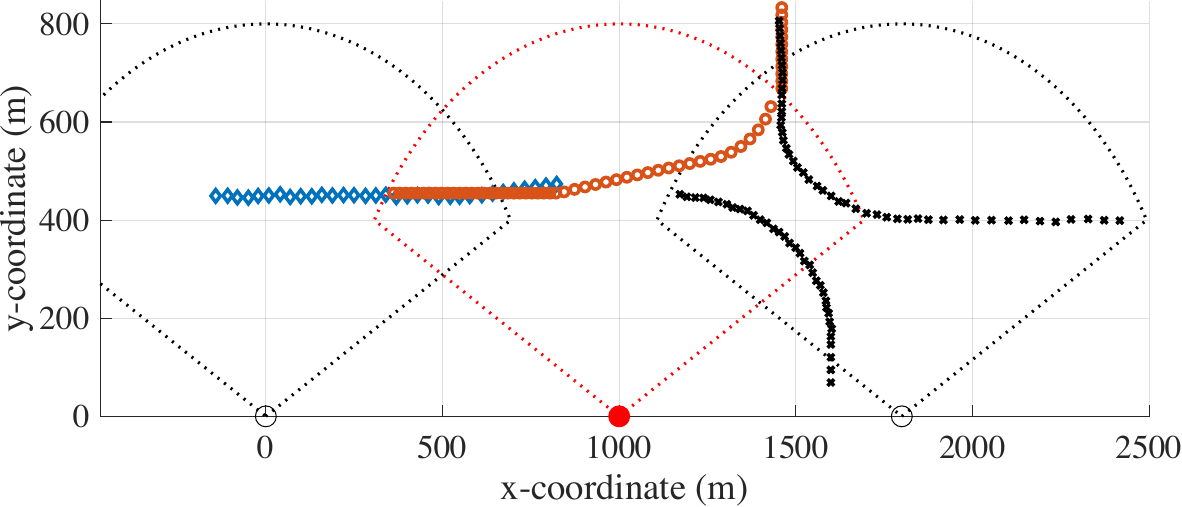}
        \caption{Individual estimates (stealthy).}\label{fig:results_1:localEst_a2}
    \end{subfigure}
    \hfill
    \begin{subfigure}[b]{0.32\textwidth}
        \centering
        \includegraphics[width=\linewidth]{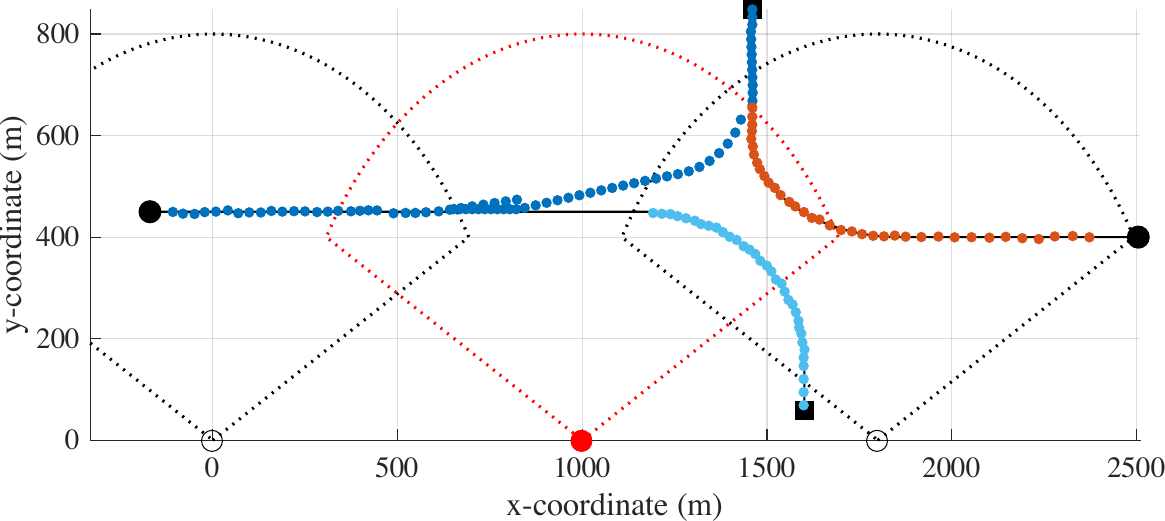}
        \caption{Consensus estimation (stealthy).}\label{fig:results_1:consensus_a2}
    \end{subfigure}
    \caption{Tracking results for one realization under hard-switch (top) and stealthy (bottom) TC-ITA attacks; the latter is generated via Algorithm~\ref{alg:FTG}. \textbf{(Left)}: Ground-truth trajectories of the victim target, the impostor target, and the spoofed trajectory.
\textbf{(Middle)}: Individual track streams available at Node~3, including its own estimate and those received from neighboring nodes, which form the inputs to the fusion stage.
\textbf{(Right)}: Local consensus estimate at Node~3, $\hat{\mathbf{X}}_{1:N}^{(3,\mathrm{con})}$.
Different colors denote different consensus label assignments; in both hard-switch and stealthy TC-ITA variants, the victim's label $\ell_\text{v}$ (blue) is successfully transferred to the impostor target.
Circle and triangle markers denote the start and end of each trajectory, respectively.}
    \label{fig:results_1}
\end{figure*}

TC-ITA relies on satisfying the track-matching constraints at the transition instants shown in Fig.~\ref{fig:attack}.
This can be achieved through a \textit{hard-switch strategy}, as illustrated in Fig.~\ref{fig:results_1:gt_a1}: once the victim label has been extracted (i.e., the spoofed track has assumed $\ell_{\mathrm{v}}$), the attacker ceases transmission during the blind interval. When the impostor target becomes visible and is reported by non-compromised nodes, the attacker resumes transmission by forwarding a trajectory that matches the impostor track $t_{\mathrm{i}}$.
While sufficient to induce label hijacking, this strategy can produce kinematically implausible trajectories that may be detected by motion-consistency checks.
To overcome this limitation, we propose a \textit{stealthy} realization of TC-ITA that enforces kinematic consistency by enabling a smooth label pull-off, as shown in Fig.~\ref{fig:results_1:gt_a2}.
\textcolor{black}{This strategy exhibits a takeover effect analogous to that observed in RGPO deception strategies~\cite{Calatrava2024DeceptionSurvey}.}


Once the condition $\tilde d^{(c)}(t_{\mathrm{v}}, t_\ast) < c$ has been satisfied in the first attack stage (Sec.~\ref{sec:stage1}), the design of the spoofed track $t_\ast$ must address three main objectives: \textit{(i)} smoothly converging toward the impostor track $t_{\mathrm{i}}$; 
\textit{(ii)} maintaining sufficient separation from the reappearing victim track $t_{\mathrm{v}}$; and 
\textit{(iii)} ensuring stealth by imposing physical constraints on the planned trajectory, such as velocity and acceleration limits ($v_\text{max}$ and $a_\text{max}$).
To systematically enforce these objectives and constraints, we formulate the design of \(t_\ast\) as a constrained optimal control problem and solve it via \gls{MPC}, which naturally accounts for system dynamics, prediction, and physical limits in trajectory generation. We emphasize that this optimization strategy is used \textit{exclusively by the attacker} to generate spoofed trajectories and is not part of the tracking or fusion process.

In the following, we use the shorthand
\( t(k) \triangleq x_k \) to associate any track \(t\) with its single-object
kinematic state at time \(k \in \mathcal{D}^{(t)}\). The state is defined as $x_k = [p_k^\top,v_k^\top]^\top \in \mathbb{X}$, where \(p_k, v_k \in \mathbb{R}^2\) denote the position
and velocity vectors.
This convention applies to all tracks introduced in the previous subsection, i.e.,  \( t_\ast(k) \triangleq x_k^\ast \), \( t_\text{i}(k) \triangleq x_{k}^\text{i} \) and \( t_\text{v}(k) \triangleq x_{k}^{\text{v}} \).




The spoofed state dynamics are modeled using a constant-velocity transition model, augmented with an attacker-controlled acceleration input, as
\begin{align}\label{eq:system_dynamics}
    x_{k}^\ast = A x_{k-1}^\ast + B u_{k-1}^\ast ,
\end{align}
\begin{equation}
A =
\begin{bmatrix}
1 & 0 & \Delta t & 0\\
0 & 1 & 0 & \Delta t\\
0 & 0 & 1 & 0\\
0 & 0 & 0 & 1
\end{bmatrix},
\qquad
B =
\begin{bmatrix}
\frac{1}{2}\Delta t^2 & 0\\
0 & \frac{1}{2}\Delta t^2\\
\Delta t & 0\\
0 & \Delta t
\end{bmatrix},
\end{equation}
where \( u_k^\ast \in \mathbb{R}^2 \) is the attacker-controlled acceleration input, and \( \Delta t > 0 \) denotes the sampling period.

%
At each time step $k_0'$, the attacker solves a $K$-step finite-horizon optimal control problem. 
The associated \emph{prediction-horizon index set} is defined as
\begin{equation}
\mathcal{K}_{k_0'} \triangleq \{k_0' + \kappa : \kappa = 0, 1, \ldots, K\}
= \{k_0', k_0'+1, \ldots, k_0'+K\},
\end{equation}
where $\mathcal{K}^{-}_{k_0'} \triangleq \mathcal{K}_{k_0'} \setminus \{k_0'+K\}$.
Under the initial condition $\bar{x}_{k_0'}$, we solve the control (acceleration) and predicted state sequences,
\begin{equation}
U^{\ast} \triangleq \big[u^{\ast}_{k}\big]_{k \in \mathcal{K}^{-}_{k_0'}}\in \mathbb{R}^{2\times K},
\quad
X^{\ast} \triangleq \big[x^{\ast}_{k}\big]_{k \in \mathcal{K}_{k_0'}}\in \mathbb{R}^{2\times (K+1)},
\end{equation}
according to the optimization problem:
\begin{equation}\label{eq:MPC_opti}
\begin{aligned}
\hat{U}^\ast, \hat{X}^\ast
&= \argmin_{U^\ast, X^\ast}
  \alpha_p \sum_{k=k_0'+1}^{k_0'+K} \gamma_p^k \|p_k^\ast - p_k^{\text{i}}\|^2 \\
&\quad + \alpha_v \sum_{k=k_0'+1}^{k_0'+K} \gamma_v^k \|v_k^\ast - v_k^{\text{i}}\|^2 \\
&\quad + \alpha_c \sum_{k=k_0'+1}^{k_0'+K}
   \max\bigl(c - \|p_k^\ast - p_k^{v}\|, 0\bigr)^2 \\
\text{s.t.}\quad
& x_{k}^\ast = A x_{k-1}^\ast + B u_{k-1}^\ast,
  \qquad k \in \mathcal{K}_{k_0'} \setminus \{k_0'\} ,\\
& x_{k_0'}^\ast = \bar{x}_{k_0'},\\
& \|u_k^\ast\| \le a_{\max},
  \qquad \forall k \in \mathcal{K}^-_{k_0'} ,\\
& \|v_k^\ast\| \le v_{\max},
  \qquad \forall k \in \mathcal{K}_{k_0'}.
\end{aligned}
\end{equation}
The first term in this objective drives the
spoofed trajectory toward the impostor target, the second enforces kinematic
consistency by penalizing velocity mismatch, and the third encodes the deception
requirement by maintaining a minimum distance from the victim target. These terms
are weighted by $\alpha_p,\alpha_v,\alpha_c \ge 0$, which balance the relative
importance of tracking, smoothness, and separation, while
$\gamma_p,\gamma_v \in (0,1]$ are exponential discount factors that prioritize
near-term deviations over those further in the prediction horizon.


Note that the same MPC structure in~\eqref{eq:MPC_opti} can be used to generate the full spoofed trajectory by adapting the reference across stages: during mimicry, the reference is set to the victim state with $\alpha_c = 0$; during pull-off, it is set to a prescribed \textit{rendezvous point}; and during injection, it is set to the current impostor state.

The MPC problem in~\eqref{eq:MPC_opti} is solved online at each time step by
predicting the system evolution over a $K$-step horizon and optimizing the
corresponding control sequence. Over a total duration of $N$ time steps, the
resulting online spoofed-trajectory generation strategy is summarized in
Algorithm~\ref{alg:FTG}.
At time $k_0'$, the current impostor and victim states,
$x_{k_0'}^{\mathrm{i}}$ and $x_{k_0'}^{\mathrm{v}}$, are obtained from the network
(e.g., from compromised or non-compromised nodes accessible to the attacker).
To construct the $K$-step trajectories required in~\eqref{eq:MPC_opti}, these states
are then propagated forward using the assumed motion model (e.g., constant
velocity), yielding auxiliary predictions that are used solely to compute the MPC
solution. At the next time step $k_0'+1$, the prediction window is shifted and the
process is repeated, with the initial states again refreshed from the latest
network track estimates via \texttt{GetFromNetwork}.

\begin{algorithm}[t]
\caption{\texttt{SpoofTrack()}}
\label{alg:FTG}
\KwIn{Initial spoofed state $\bar{x}_{0}$, initial impostor state $x_{0}^{\mathrm{i}}$,
initial victim state $x_{0}^{\mathrm{v}}$, prediction horizon $K$, attack duration $N$,
OSPA threshold $c$, velocity and acceleration bounds $v_{\max}$, $a_{\max}$.}
\KwOut{Spoofed state sequence $\mathcal{S}$}

\textbf{Initialization:} Record $x_{0}^\ast \gets \bar{x}_{0}$ in $\mathcal{S}$; \\
\For{$n \gets 0$ \textbf{to} $\mathrm{N}-1$}{
$(x_{n}^{\mathrm{i}}, x_{n}^{\mathrm{v}}) \gets \texttt{GetFromNetwork}(n)
$\;
$\hat{U}^{\ast} \gets$
Solve~\eqref{eq:MPC_opti} for $k_0' \gets n$ and parameters \\ \qquad \qquad 
$\{x_{n}^\ast, x_{n}^{\text{i}}, x_{n}^{\text{v}}, \mathrm{K}, c , a_{\textrm{max}}, v_{\textrm{max}}\}$\;
$u_{n}^\ast \gets \hat{U}^\ast[:,1]$ Get first optimal control;
%
$x_{n+1}^\ast \gets Ax_{n}^\ast + Bu_{n}^\ast$ Update spoofed state}
\Return $\mathcal{S} = [x_{0}^\ast, \ldots, x_{N}^\ast]$
\end{algorithm}


While the trajectory-alignment terms towards $t^\text{i}$ in~\eqref{eq:MPC_opti} are convex and can be solved using standard quadratic optimization tools, the separation penalty introduces a nonconvex component.
By formulating this requirement as a soft penalty, the optimization remains feasible while allowing a controlled trade-off between trajectory fidelity and separation from the true target. This is particularly important in scenarios where the spoofed and true trajectories are initially close and strict avoidance may be temporarily infeasible. In our simulations, the resulting nonlinear program is solved using \textit{CasADi}~\cite{andersson2019}.

\section{Simulation Experiments}\label{sec:results}

%

This section evaluates the hard-switch and stealthy TC-ITA variants introduced in Sec.~\ref{sec:attack:mpc}.
We illustrate label hijacking in a representative three-node case study and assess its impact on tracking performance.

\subsection{Simulation Setup}

We consider a three-node distributed sensor network tracking two physical targets over a
$[-500, 2500]\,\mathrm{m} \times [0, 1000]\,\mathrm{m}$ area: a victim target, whose identity is hijacked, and an impostor target, to which the identity is transferred.
The sensor nodes are located at $[0,0]^\top$, $[1000,0]^\top$, and $[1800,0]^\top$\,m and are indexed from left to right as Nodes~1--3, with only Node~2 compromised by the attacker.

We implement the proposed TC-ITA strategy within the TC-DMTT framework of~\cite{van2021distributed}, which employs a local \gls{LMB} filter at each node.
A constant-velocity motion model is assumed, with process noise standard deviation $\sigma_v = 5$ and measurement noise standard deviation $\sigma_r = 2\,\mathrm{m}$.
Each sensor detects objects within a relative angle of $[-60^\circ, 60^\circ]$, with detection probability $P_D = 0.98$ and detection range $r_D = 800\,\mathrm{m}$.
Consistent with~\cite{van2021distributed}, the base distance metric in~\eqref{eq:ospa_track2track} is chosen as the Manhattan distance, i.e., $d(x,y)=\|x-y\|_1$.
%

%

The target ground-truth trajectories over the evaluation window $N = 80$ are depicted in Figs.~\ref{fig:results_1:gt_a1} and~\ref{fig:results_1:gt_a2}, where the physical targets $t_{\mathrm{v}}$ and $t_{\mathrm{i}}$ appear in green and blue, respectively, and the spoofed track $t_\ast$ appears in red for both attacks under evaluation.
The stealthy TC-ITA trajectory in Fig.~\ref{fig:results_1:gt_a2} is generated using the MPC-based strategy summarized in Algorithm~\ref{alg:FTG}, with horizon $K = 20$, weights $\alpha_p = 1$, $\alpha_v = \alpha_c = 0.1$, discount factors $\gamma_p = \gamma_v = 0.99$, and physical bounds $v_{\max} = 30\,\mathrm{m/s}$ and $a_{\max} = 30\,\mathrm{m/s^2}$.
%

Figure~\ref{fig:results_1} shows results for a single representative 
realization, while Figs.~\ref{fig:cardinality_Est} and~\ref{fig:ecdf} 
report statistics averaged over $M = 100$ Monte Carlo runs.


%


\subsection{Label Hijacking Performance}


We report local fusion (consensus) results for Node~3, i.e., $\hat{\mathbf{X}}_{1:N}^{(3,\mathrm{con})}$.
This is without loss of generality, since equiprobable weights are used in the kinematic consensus stage (Sec.~\ref{sec:kin_consensus}).

As evidenced by target and spoofed trajectories in 
Figs.~\ref{fig:results_1:gt_a1} and~\ref{fig:results_1:gt_a2}, 
the hard-switch attack produces an abrupt and kinematically 
implausible trajectory, whereas the MPC-based stealthy attack 
generates a smooth pull-off that closely follows the impostor 
target while maintaining separation from the victim, resulting 
in a visually plausible spoofed trajectory $t_*$.

The corresponding individual track streams available at Node~3 (namely its own local estimate $\hat{\mathbf{X}}_{1:N}^{(3)}$ and those received from its neighbors $\{\hat{\mathbf{X}}_{1:N}^{(i)}\}_{i\in\mathcal{N}(3)}$) are shown in Figs.~\ref{fig:results_1:localEst_a1} and~\ref{fig:results_1:localEst_a2}.
These track streams constitute the inputs to the fusion stage, 
and show that the spoofed track $t_*$, injected by the 
compromised node (Node~2, orange), enters the fusion process 
as a legitimate track estimate, indistinguishable from those 
reported by honest nodes.

Consensus results are shown in 
Figs.~\ref{fig:results_1:consensus_a1} 
and~\ref{fig:results_1:consensus_a2}, where different colors 
correspond to different consensus label assignments.
The victim 
target is initially associated with the blue label, while the 
impostor carries the red label; both attack variants 
successfully transfer the blue label to the impostor.
The 
hard-switch strategy achieves this through an abrupt trajectory 
transition that, while kinematically implausible, remains 
unmitigated by the TC-DMTT framework.
This highlights that TC-DMTT frameworks rely solely on 
metric-based matching for label consensus, overlooking 
kinematic consistency as an additional integrity check.
The stealthy TC-ITA further evades potential motion-consistency checks through its dynamically plausible trajectory.

\begin{figure}[t]
\centering
\subfloat[Mean estimated cardinality.\label{fig:cardinality_Est}]{%
    \includegraphics[width=.8\columnwidth]{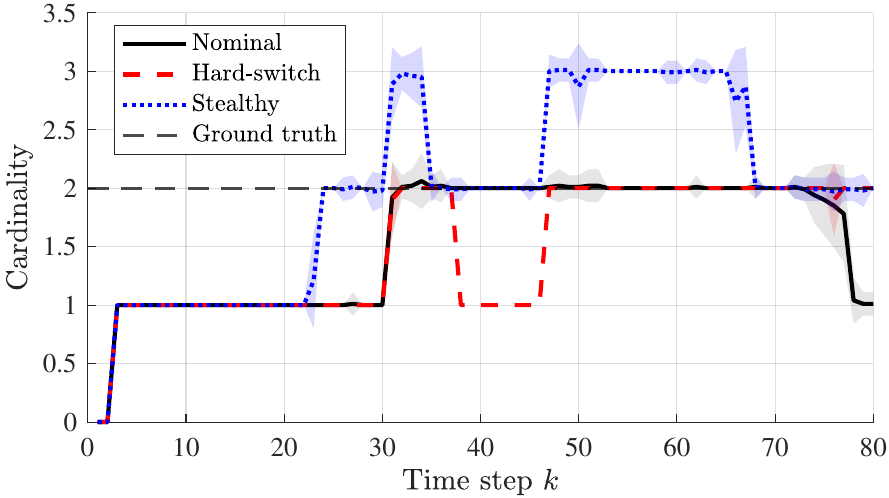}}\\
\subfloat[ECDF of OSPA(2).\label{fig:ecdf}]{%
    \includegraphics[width=.75\columnwidth]{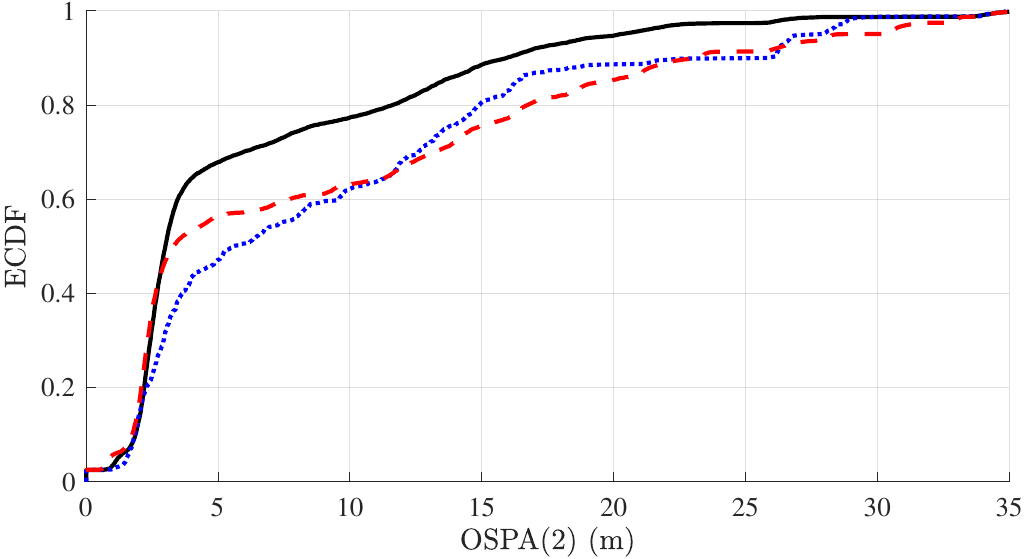}}
\caption{Impact on tracking accuracy over the full evaluation window at Node~3 under hard-switch and stealthy TC-ITA, and no attack (nominal scenario).}
\label{fig:tracking_accuracy}
\end{figure}

Cardinality estimation results are shown in Fig.~\ref{fig:cardinality_Est}.
For the hard-switch attack, a transient underestimation 
occurs mostly during the switching interval, when the attacker ceases transmission.
In contrast, under the stealthy TC-ITA, the cardinality is 
overestimated, as the spoofed track $t_*$ does not 
necessarily coincide with any true target throughout the entire attack.
%
%
Fig.~\ref{fig:ecdf} shows the empirical CDF (ECDF) of the 
instantaneous OSPA across Monte Carlo runs for hard-switch and stealthy TC-ITA, and the nominal scenario (no attack). Both attack 
variants significantly degrade tracking accuracy relative 
to the nominal case, confirming that label hijacking also 
impacts target tracking performance.

%



\section{Conclusion}\label{sec:conclusion}

This paper reveals identity-level vulnerabilities in track-consensus
distributed multi-target tracking.
By exploiting the metric-based track-matching stage, the proposed
track-consensus identity-theft attack enables label hijacking and
target impersonation.
We introduce a hard-switch attack realization, showing that even abrupt and
kinematically implausible trajectories suffice to trigger label hijacking under
OSPA-based matching.
We then propose a stealthy MPC-based realization that enforces smooth,
physically consistent pull-off behavior while satisfying the same matching
constraints.
Simulations in a limited-FoV network confirm that both variants succeed.
These findings highlight the need for identity-aware protection at the
consensus layer.

%
%

\bibliographystyle{ieeetr}%
\bibliography{tracking}

\end{document}